\documentclass[a4paper,prl,final,superscriptaddress,twocolumn]{revtex4}
\usepackage[latin1]{inputenc}
\usepackage[T1]{fontenc}
\usepackage[english]{babel}
\usepackage{graphicx}
\usepackage{amsmath}
\usepackage{amssymb}
\usepackage{mathrsfs}
\usepackage{textcomp}
\usepackage{tabularx}
\usepackage{color}
\usepackage{array}

\begin{document}

\title{Manipulating the magnetic state of a carbon nanotube Josephson junction using the superconducting 
phase}
\author{R. Delagrange}
\affiliation{Laboratoire de Physique des Solides, Univ. Paris-Sud, CNRS, UMR 8502, F-91405 Orsay Cedex, France.}
\author{D. J. Luitz} 
\affiliation{Laboratoire de Physique Th\'eorique, IRSAMC, Universit\'e de Toulouse and CNRS, 31062
Toulouse, France}
\author{R. Weil}
\affiliation{Laboratoire de Physique des Solides, Univ. Paris-Sud, CNRS, UMR 8502, F-91405 Orsay Cedex, France.}
\author{A. Kasumov}
\affiliation{Laboratoire de Physique des Solides, Univ. Paris-Sud, CNRS, UMR 8502, F-91405 Orsay Cedex, France.}
\author{V. Meden}
\affiliation{Institut f{\"u}r Theorie der Statistischen Physik, RWTH Aachen University and JARA---Fundamentals 
of Future Information Technology, 52056 Aachen, Germany} 
\author{H. Bouchiat} 
\affiliation{Laboratoire de Physique des Solides, Univ. Paris-Sud, CNRS, UMR 8502, F-91405 Orsay Cedex, France.}
\author{R. Deblock}
\affiliation{Laboratoire de Physique des Solides, Univ. Paris-Sud, CNRS, UMR 8502, F-91405 Orsay Cedex, France.}

\begin{abstract}

The magnetic state of a quantum dot attached to superconducting leads is experimentally shown to be controlled by the superconducting phase difference across the dot. This is done by probing the relation between the Josephson current and the superconducting phase difference of a carbon nanotube junction whose Kondo energy and superconducting gap are of comparable size. It exhibits distinctively anharmonic behavior, revealing a phase mediated singlet to doublet transition. We obtain an excellent quantitative agreement with numerically exact quantum Monte Carlo calculations. This provides strong support that we indeed observed the finite temperature signatures of the phase controlled zero temperature level-crossing transition originating from strong local electronic correlations.

PACS number(s): 74.50.+r, 72.15.Qm, 73.21.-b, 73.63.Fg 

\end{abstract}

\maketitle

When a localized magnetic moment interacts with a Fermi sea of conduction electrons, the Kondo effect 
can develop: spin-flip processes lead to a many-body singlet state in which the 
delocalized electrons screen the moment. Quantum dots (QD) in the Coulomb blockade regime 
and particularly carbon nanotube (CNT) dots constitute ideal systems for the investigation of Kondo 
physics at the single spin level \cite{Pustilnik2004,Goldhaber-Gordon1998,Cronenwett1998}. 
In these systems, it is possible to control the number of electrons on the dot varying a gate voltage. For an odd occupation, the dot accommodates a magnetic moment which is screened provided that this is not prohibited by an energy scale larger than the 
Kondo energy $k_B T_K$. Temperature is the most obvious obstacle to the development of the Kondo effect since $T_K$ can be smaller than 1K. However, if temperature is sufficiently low, the Kondo  effect may compete with other quantum many-body phenomena such as superconductivity, for which the formation of 
Cooper pairs of energy $\Delta$ may prevent the screening of the dot's spin. This situation can be 
investigated using superconducting hybrid junctions, where a supercurrent is induced by the 
proximity effect, for example in CNT-based QDs \cite{Kasumov1999} or semiconductor-based ones \cite{Doh2005}.

A setup of a high resistance tunnel barrier between two superconductors, also called Josephson junction (JJ), carries a  supercurrent $I=I_C  \sin \varphi$, with the critical current $I_C$. The superconducting phase difference across the junction $\varphi$ controls the amplitude and the sign of the supercurrent. This is the Josephson relation, the most famous example of a current-phase relation (CPR). In some peculiar systems such as ferromagnetic superconducting junctions, the transmission of Cooper pairs gives rise to a $\pi$ phase shift of the CPR \cite{Ryazanov2001}. In QD JJs (tunnel barrier replaced by QD) in the strong Coulomb blockade regime where the Kondo effect is negligible,  such a $\pi$ shift is observed as well since the tunneling of a Cooper pair implies reversing the order of particles within this pair. This leads to a gate-controlled sign reversal of the CPR when the parity of the number of electrons is changed, as was observed experimentally  \cite{Vandam2006,cleuziou2006,Jorgensen2007,defranceschi2010}.  
In contrast, if the Kondo effect and thus local correlations prevail, the spin of the dot is screened by unpaired electrons leading to a singlet ground state: the 0-junction is then recovered even though the parity of the dot is still odd. 

\begin{figure}[ht]
                \begin{center}
                \includegraphics[height=6cm]{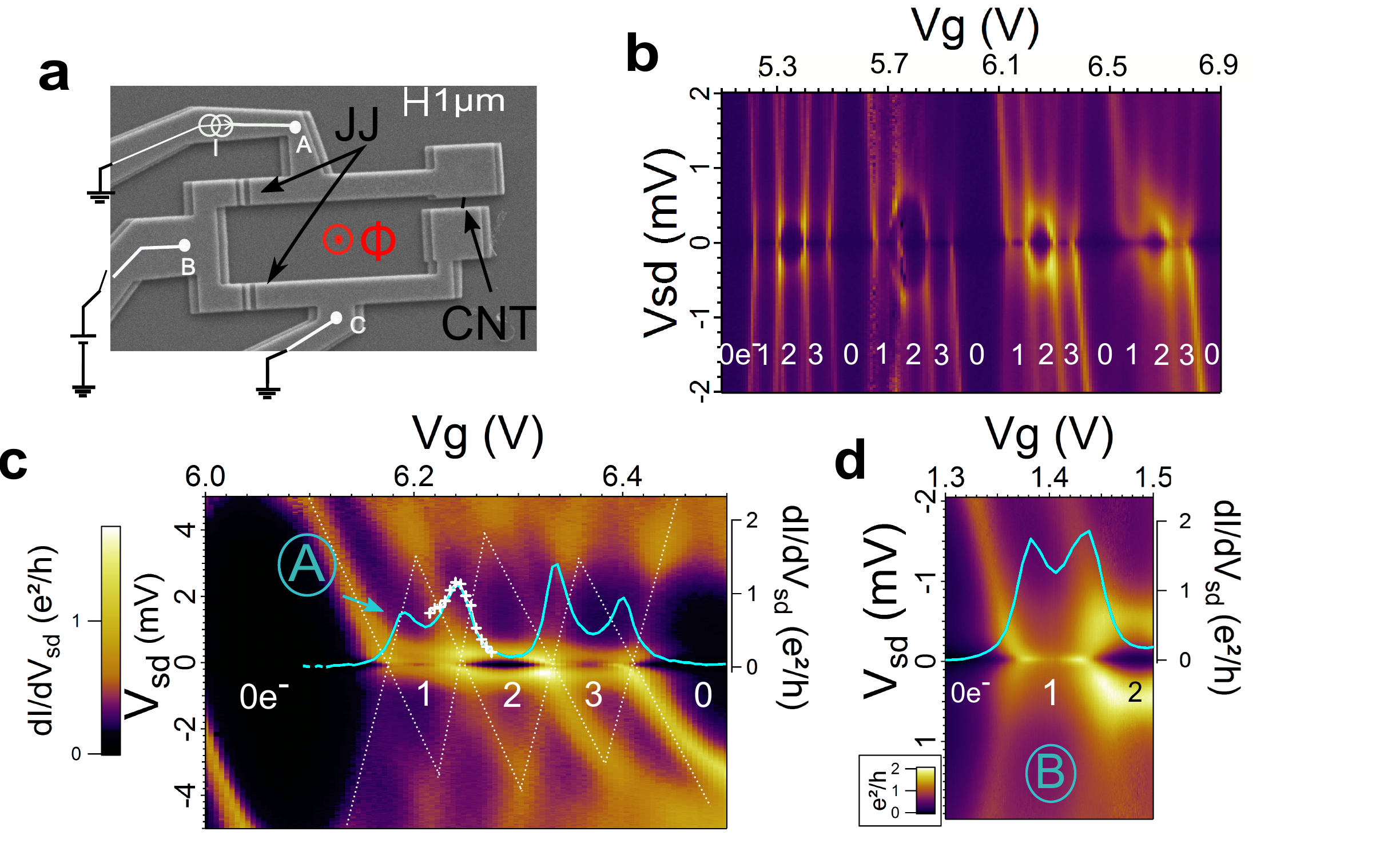}
                \end{center}
                \caption{ \textbf{a} 
                                 Scanning electron microscopy image of the measured asymmetric SQUID (see text and \cite{Basset2014}). \textbf{b} Differential conductance ($\frac{\mathrm{d}I}{\mathrm{d}V_{sd}}$) 
                                in the normal state of the CNT junction versus gate voltage $V_g$ and bias voltage $V_{sd}$ for a large range of $V_g$. Figures \textbf{c} and \textbf{d} focus on two Kondo zones called respectively A and B. A 1T magnetic field is applied to destroy superconductivity in the contacts. The number of electrons in the last occupied energy levels is indicated in white.
                                In light blue, $\frac{\mathrm{d}I}{\mathrm{d}V_{sd}}(V_{sd})$ at zero bias is plotted  
                                (axis on the right). In the normal state, the reference JJ contribution is a constant which was subtracted to obtain the plots. 
                 The white symbols in \textbf{c} correspond to the theoretical fit of the conductance (see text).}
                \label{didv_normal}
                \end{figure} 

The switching from 0- to $\pi$-junction behavior as a function of a variety of energy scales of the QD JJ in the presence of local correlations was extensively studied theoretically 
\cite{Clerk2000,Glazman1989,Vecino2003,choi2004,Siano2004,Bauer2007,Karrasch2008,Meng2009,Luitz2010,Luitz2012}. The scales are the broadening  $\Gamma$ of the energy levels in the dot due to 
the coupling to the reservoirs, the superconducting gap $\Delta$ of the contacts, the dot's charging energy $U$ and its level energy $\epsilon$. When these parameters fall into the 0-$\pi$ transition regime, it was predicted that the ground state of the system---singlet or doublet---depends on the phase difference $\varphi$, undergoing a level-crossing transition. This leads to a characteristic anharmonicity of the CPR for temperature $T>0$ and a jump at a critical phase $\varphi_C$ for $T=0$. In other words, in this particular regime of parameters, the magnetic state of the dot is predicted to be governed by the superconducting phase difference across the junction.

This Kondo related 0-$\pi$ transition was earlier observed 
experimentally as a function of the gate voltage \cite{Jorgensen2007,Eichler2009,Maurand2012} and the spectroscopy of 
Andreev bound states enabled a better understanding of the involved 
physics \cite{Pillet2010,Pillet2013,Chang2013,Deacon2010,Kim2013}. 
Measurements of the CPR of a QD JJ embedded in a SQUID were also performed \cite{Maurand2012} and 
indeed showed anharmonicities. However, in the region of 0-$\pi$ transition, the obtained CPRs are not 
odd functions of flux as they should be, indicating that the physics is spoiled by other 
effects \cite{note_EM_environment,Avriller2015}.

Here we report on the successful measurement of the CPR of a CNT-based hybrid junction over the entire 0-$\pi$ transition. This constitutes the first experimental demonstration of the 0-$\pi$ transition controlled by the superconducting phase $\varphi$. 
A very important part of our analysis is the comparison between the measured CPRs and
theoretical ones, computed for the Anderson model with superconducting leads using a numerically exact quantum Monte Carlo (QMC) method. The excellent agreement provides strong support 
that we indeed observe the transition resulting from strong local electronic correlations.

 We fabricated a CNT-based QD, connected to superconducting leads and embedded in an asymmetric modified SQUID (Fig. \ref{didv_normal} \textbf{a}). This device, a SQUID containing the QD JJ (here the CNT) and a reference JJ with critical current high compared to the one of the QD JJ, allows us to determine the CPR of interest \cite{DellaRocca2007,Basset2014}. The switching current $I_s$ of the SQUID versus magnetic flux is measured. The CPR of the QD JJ is then obtained by extracting the modulation of $I_s$ around its mean value $\left<I_s \right>$. Our device possesses a second reference JJ and a third connection as described in Ref. \cite{Basset2014}. This allows us to characterize each junction independently at room temperature, and to measure both the CPR of the CNT and its differential conductance in the superconducting state.

The CNTs are grown by chemical vapor deposition on an oxidized doped silicon 
wafer \cite{Kasumov2007}. A three-junctions SQUID is constructed around a selected nanotube with the following materials: Pd(7~nm)/Nb(20~nm)/Al(40~nm), AlOx and Al(120~nm) \cite{supp_materials}.  The sample is thermally anchored to the mixing chamber of a dilution refrigerator of base temperature 50 mK and 
measured through low-pass filtered lines. A magnetic field $B$ is applied perpendicular to the loop to modulate the phase difference across the CNT-junction by  $2\pi B S/ \Phi_0$, with the superconducting flux quantum 
$\Phi_0=h/2e$ and S the loop area.

  We first characterize the sample in the normal state, measuring the differential conductance 
$\mathrm{d}I/\mathrm{d}V_{sd}$ 
versus bias voltage $V_{sd}$ for various backgate voltages $V_g$, using a lock-in-amplifier technique. 
The contacts are made of Pd/Nb/Al with a gap of $\Delta=0.17~\mathrm{m eV}\pm 10 \%$, a value very close 
to the gap of Al but considerably smaller than the Nb gap because of the Pd layer. A magnetic field of 1T 
is needed to suppress superconductivity in these contacts. Even though such a magnetic field significantly 
affects the Kondo effect, the results of Fig. \ref{didv_normal} show Coulomb diamonds and an increase of the conductance at zero-bias in some diamonds, a signature of the Kondo effect. 
The four fold degeneracy, characteristics of clean carbon nanotubes with orbital degeneracy \cite{Cleuziou2013, Schmid2015}, is clearly seen (Fig. \ref{didv_normal} \textbf{b}). This allows us to determine the dot's occupancy indicated on the figure.
We focused on two ranges of gate voltages, corresponding to diamonds with odd occupancy, where non-zero conductance is observed at zero bias, zone A (around $V_g=6.2~\mathrm{V}$) and zone B (around $V_g=1.4~\mathrm{V}$) (see Figs. 1 \textbf{c} and \textbf{d}). Among all the 
Kondo ridges leading to a $\pi$-junction in the superconducting state, those show the widest extent in gate voltage of the 0-$\pi$ transition, which makes the measurements more accurate. The height in $V_{sd}$ of the Coulomb diamonds gives the charging energy $U=3.2~\mathrm{meV}\pm 10\%$ in zone A and $2.5~\mathrm{meV}\pm10\%$ in zone B. 
Due to the complex interplay of the Kondo scale $k_B T_K$ and the Zeeman energy, that are of comparable size, $\Gamma$ and the contact asymmetry cannot be determined directly from the experimental results; theoretical modeling is required. 
  
The junction is modeled by an Anderson impurity model \cite{Clerk2000,Glazman1989,Vecino2003,choi2004,Siano2004,Bauer2007,Karrasch2008,Meng2009,Luitz2010,Luitz2012} with right ($R$) and left ($L$) BCS superconducting 
leads and superconducting order parameter $\mathrm{e}^{\pm \mathrm{i}\varphi/2}\Delta$. The interaction 
of electrons on the QD is given by a standard 
Hubbard term with charging energy  $U$ and the coupling of the leads to the QD is described by the 
energy independent hybridization strength $\Gamma_{L/R}$. We solve it using the numerically exact CT-INT Monte Carlo method \cite{Luitz2010} in the normal 
state ($\Delta=0$) in a magnetic field and calculate the finite temperature linear conductance for different 
$\Gamma_{L/R}$ as a function of the dot energy $\epsilon$, defined relative to particle-hole symmetry. 
The amplitudes of the magnetic field and the charging energy are fixed to the experimentally determined 
values $B=1 \, \mathrm{T}$ and $U=3.2\, \mathrm{meV}$ of zone A \cite{numerics_costs}. A comparison with the 
measured conductance at zero source drain voltage (i.e. in equilibrium) yields a set of parameters that 
fit the experiment best. We 
find $\Gamma_R+\Gamma_L = 0.44\, \mathrm{meV}$, $\Gamma_R/\Gamma_L=4$. We also slightly varied $T$ to 
estimate the electronic temperature in the sample and obtain $T=150\, \mathrm{mK}$. Additionally, 
this procedure gives a reliable way of extracting the conversion factor $\alpha=39\,\frac{\mathrm{meV}}{\mathrm{V}}$ 
between the applied gate voltage and the dot on-site energy $\epsilon$ (as done in \cite{Luitz2012}). Our best fit is displayed in 
Fig. \ref{didv_normal} \textbf{c} (white symbols). 
After reliably estimating all parameters, using the formula $T_K=\sqrt{\Gamma U/2}\exp(-\pi\frac{|4\epsilon^2-U^2|}{8\Gamma U})$ \cite{Haldane1978}, we can directly show that for zone A the dot is indeed in the regime of strongest competition between Kondo correlations and superconductivity with $k_B T_K \approx \Delta$ (Fig. \ref{analyse} \textbf{a}).

  \begin{figure}[ht]
                  \begin{center}
                  \includegraphics[width=\columnwidth]{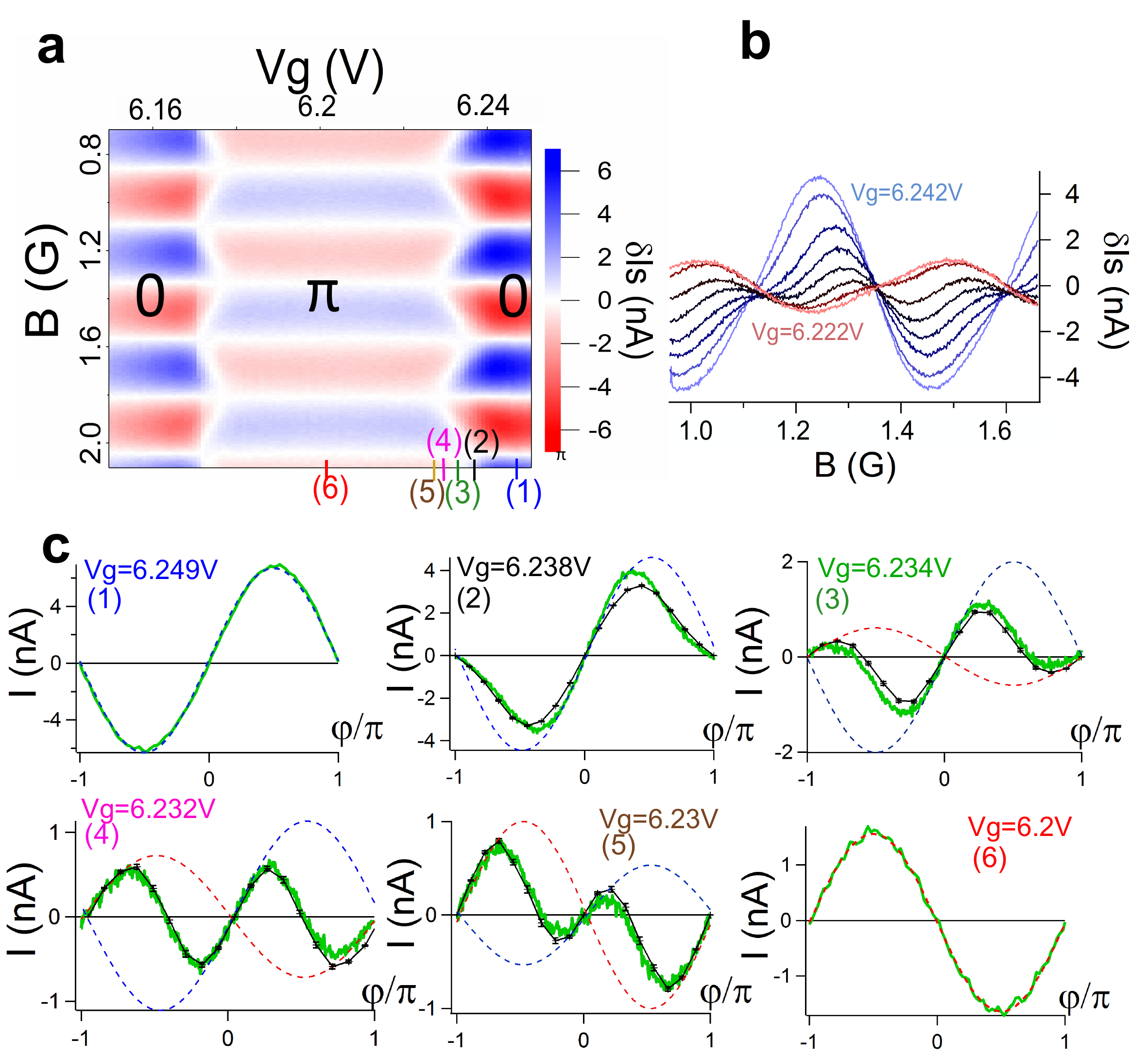}
                  \end{center}
                  \caption{\textbf{a} Modulation of the switching current of the SQUID, proportional to the CPR of the CNT junction, versus the magnetic field for various gate voltages in zone A. \textbf{b} 
                     Modulation of the switching currents near the transition for several gate voltages. \textbf{c} CPR extracted from the previous data and rescaled by 1.33 (see the text) near the transition (green continuous line). The theoretical predictions resulting from QMC calculations are shown as black lines. The dashed lines are guides to the eyes and represent the contributions of the singlet (0-junction, in blue) and the doublet state ($\pi$-junction, in red).}
                  \label{Is_A}
                  \end{figure}  
  
   \begin{figure}[ht]
          \begin{center}
          \includegraphics[width=\columnwidth]{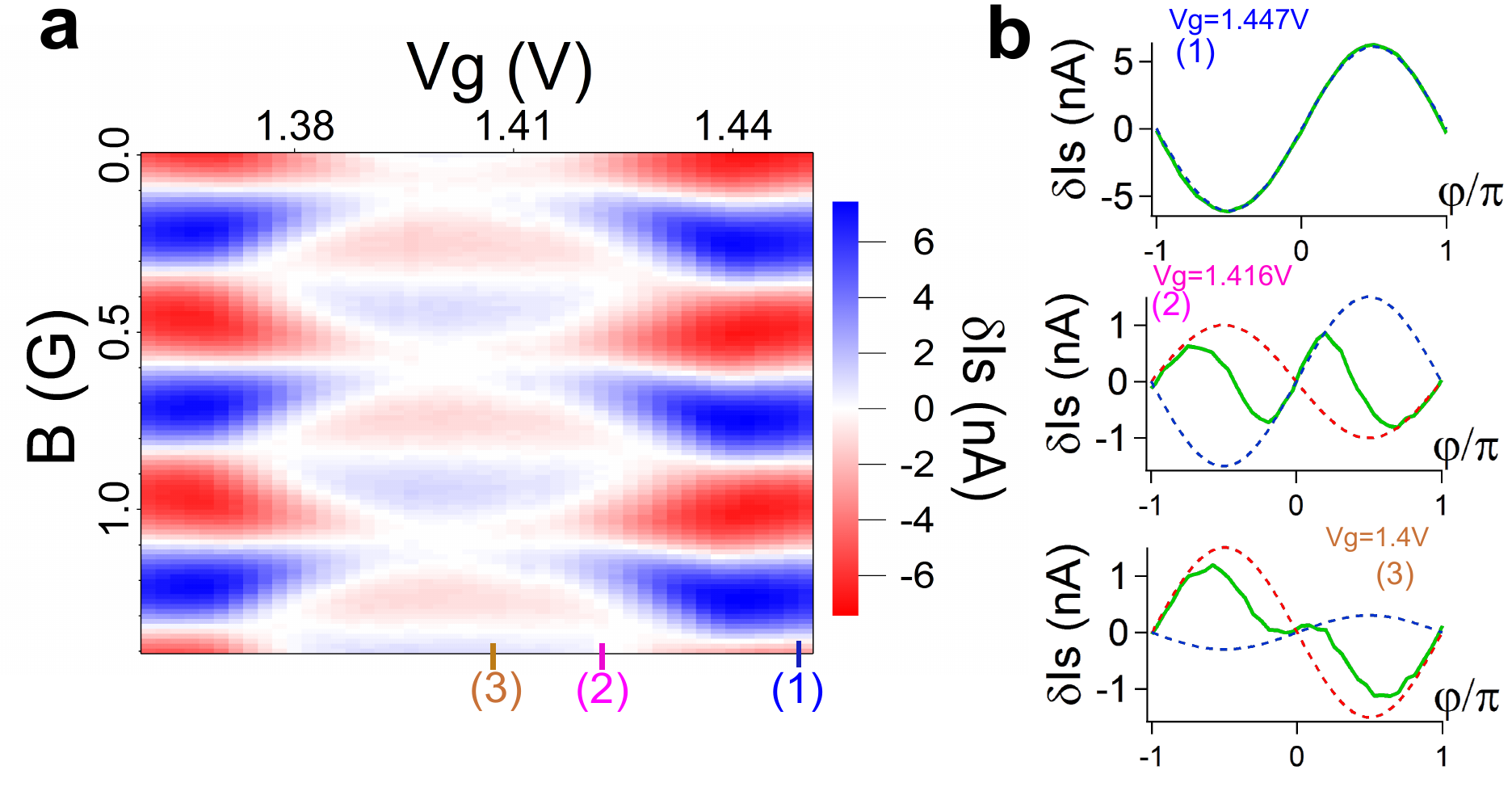}
          \end{center}
          \caption{\textbf{a} Modulation of the switching current of the SQUID for zone B, which exhibits a partial transition from 0- to $\pi$-junction. \textbf{b} CPR extracted from the previous data near the transition (green continuous line). The dashed lines are guides to the eyes and represent the contributions of the singlet (0-junction in blue) and the doublet state ($\pi$-junction in red).}
          \label{Is_B}
          \end{figure}
 
Next, superconductivity is restored by suppressing the 1T magnetic field and the CPR is measured in both Kondo zones, extracting the modulation of the switching current $\delta I_s$ versus magnetic field (Fig. \ref{Is_A}) from the critical current of the SQUID. To measure the switching current, the SQUID is biased with a linearly increasing current with a rate  $\frac{\mathrm{d}I}{\mathrm{d}t}=37~\mathrm{\mu A/s}$ and the time at which the SQUID switches to a dissipative state is measured. This process is reproduced and averaged around 1000 times, the whole procedure being repeated at different values of magnetic field below a few Gauss, small enough to preserve superconductivity. To obtain the modulation of the switching current $\delta I_s$ versus magnetic field, the contribution of the reference junctions (around $90~\mathrm{nA}$) is subtracted.
As demonstrated in Ref. \cite{Basset2014}, $\delta I_s$ is proportional to the CPR of the CNT junction. It should be noted that this kind of system, in particular near the 0-$\pi$ transition, is very sensitive to the electromagnetic environment, which therefore needs to be optimized \cite{note_EM_environment}.

The main results of this work are presented in Figs. \ref{Is_A} and \ref{Is_B}, where we 
show the extracted CPRs for gate values over the entire 
transition regime (curves \textbf{c}.1 to \textbf{c}.6 of Fig. \ref{Is_A} for zone A and 
curves \textbf{b}.1 to \textbf{b}.3 of Fig. \ref{Is_B} for zone B). We now analyze qualitatively the shape of these curves. 

On the edges of Kondo zone A, far from the transition (Fig. \ref{Is_A}  \textbf{c}.1), the junction  behaves as a regular 
JJ with a CPR proportional to $\sin (\varphi)$ (0-junction). In contrast, at the center of the Kondo zone 
(Fig. \ref{Is_A} \textbf{c}.6) the CPR is $\pi$-shifted ($\delta I \propto \sin(\varphi + \pi)$) and has a smaller 
amplitude characteristic for a $\pi$-junction. In between, the CPR is composite with one part corresponding 
to 0- and another part to $\pi$-junction behavior. The latter first occurs around 
$\varphi=\pi$, giving rise to a very anharmonic CPR. In the middle of the transition region, we find period 
halving (Fig. \ref{Is_A} \textbf{c}.4) \cite{halving_period}. 
This evolution of the CPR between a 0- and $\pi$-junction is consistent with the 
finite temperature transition of the dot's magnetic state (between singlet and doublet) which is controlled 
by the superconducting phase difference \cite{Glazman1989,Karrasch2008,Luitz2010}. 

A more precise analysis of the transition allows to attribute this 0-$\pi$ transition to a competition between 
the Kondo effect and the superconductivity. Indeed, around the center of Kondo zone A (Fig. \ref{Is_A} \textbf{a}), the $\pi$-junction extends over a range of $60~\mathrm{mV}$ of gate voltage. According to the $\mathrm{d}I/\mathrm{d}V_{sd}$ in the normal 
state (Fig. \ref{didv_normal}) and the conversion factor $\alpha$, the odd diamond has a width in gate voltage of 
about $82~\mathrm{mV}$, larger than the $\pi$-junction regime. Consequently, this 0 to $\pi$ transition 
is not simply due to a change in the parity of the dot filling but to an increase in the ratio $\Delta/T_K$ (see Fig. \ref{analyse} \textbf{a}). This is even more obvious for Kondo 
zone B (Fig. \ref{Is_B} \textbf{b}) where the 0 to $\pi$ transition is incomplete.

For a quantitative comparison between theory and experiment, we performed a second CT-INT calculation in 
the superconducting state ($B=0$) for zone A \cite{numerics_costs} to obtain the CPRs in the transition regime. 
We used the measured value of the superconducting gap $\Delta=0.17\,\mathrm{meV}$ and the previously determined 
parameters and computed the Josephson current as a function of the phase difference $\varphi$. The theoretical CPR 
are calculated at various $\epsilon$ (related to $V_g$ by $\epsilon=\alpha V_g$) and plotted as black lines in 
comparison to our experiments in Fig. \ref{Is_A} \textbf{c}2 to \textbf{c}5. Since our setup yields a
switching current that is
necessarily smaller than the supercurrent, the experimental CPRs were multiplied by a unique correction factor chosen 
to obtain the best agreement with the QMC results. The agreement for the shape of the CPR is excellent; however 
a shift of the energy level $\delta\epsilon=0.28~\mathrm{meV}$ of the theoretical CPRs is needed to superimpose 
them with the experimental ones \cite{supp_materials}. The QMC calculations predict a transition region centered around a smaller 
$\epsilon$ than measured experimentally (see supplementary materials); a deviation between experiment and theory 
which we currently do not understand.
Note however that the width of this transition is very well reproduced.  

   \begin{figure}[h]
         \begin{center}
                 \includegraphics[width=\columnwidth]{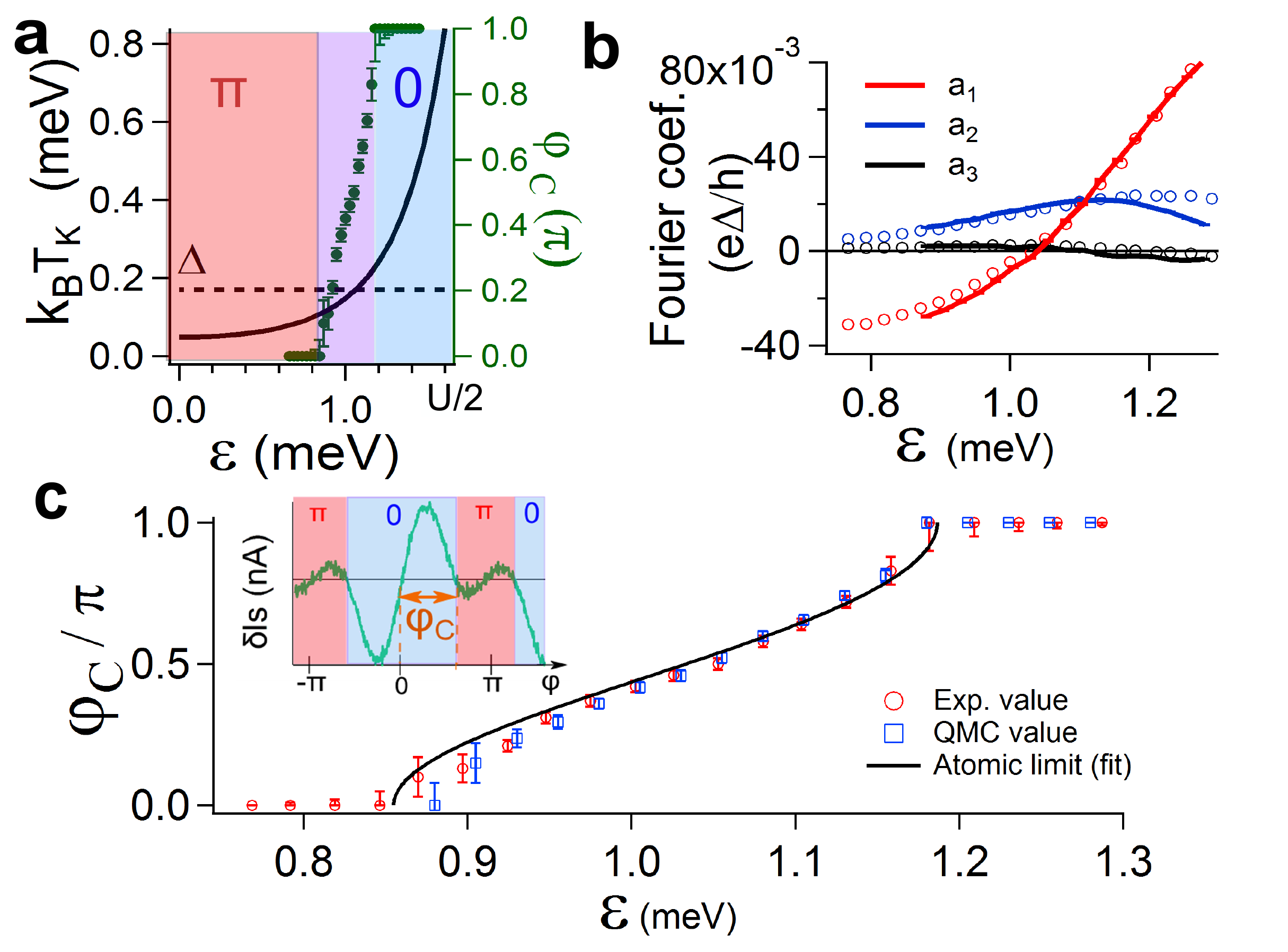}
         \end{center}
         \caption{
         \textbf{a} Calculated $T_K$ (black line) is compared to $\Delta$ (dotted line) as a function of the level energy $\epsilon$ ($\epsilon=\alpha \delta V_g$ with $\delta V_g$ the gate value measured from half-filling and $\alpha=39~\mathrm{meV/V}$). In green dots, the critical phase $\varphi_C$ is plotted, defined as the phase, different from 0 and $\pi$, for which the CPR equals zero (see text and inset of \textbf{c}). The 0-$\pi$ transition occurring when $\varphi_C$ switches from 1 to 0, the system is indeed in the regime $k_B T_K \approx \Delta$. \textbf{b} Fourier analysis of the CPRs of zone A near the transition for different level energies $\epsilon$. The circles correspond to the experimental CPRs whereas the continuous lines correspond to the QMC calculation (red, blue and black : harmonics 1, 2 and 3). \textbf{c} Measured $\varphi_C$ (red circles) versus $\epsilon$. $\varphi_C$ extracted from the QMC calculation and shifted is also shown (blue squares). The black line shows the result of a two-parameter fit of the analytical curve obtained in the atomic limit (see text).}
         
         \label{analyse}
         \end{figure}

The comparison of the measurements and calculations can even be refined employing Fourier decompositions 
$I(\varphi)= a_1 \sin(\varphi) + a_2\sin(2\varphi) +a_3\sin(3\varphi)+\dots$ of the $2 \pi$-periodic CPRs. 
The first three amplitudes suffice to describe the experiment perfectly; see Fig. \ref{analyse} \textbf{b} where they 
are  shown as functions of $\epsilon$. The theoretical model thus exactly captures the nontrivial finite 
temperature phase dependence of the measured Josephson current.

An important information that can also be extracted from the experiment is the gate voltage dependence of 
the critical phase $\varphi_C$ at which the system switches from 0 to $\pi$-junction behavior, \textit{i.e.} 
the CPR has 0-behavior for $\varphi \in [0,\varphi_C]$ and $\pi$-behavior for $\varphi \in [\varphi_C,2\pi-\varphi_C]$ 
(Fig. \ref{analyse} \textbf{c}). At $T=0$ this switching at $\varphi_C$ is associated to a first order level crossing 
transition and appears as a jump of the current from positive to negative. For $T>0$ the transition is washed out 
and the CPR is smoothed. However, at small enough $T$, $\varphi_C$ depends only weakly on $T$ (\textit{cf.} 
Refs.\cite{Karrasch2008,Luitz2010,supp_materials}). In Fig. \ref{analyse} \textbf{c}, 
we compare $\varphi_C(\epsilon)$ from experiment and theory. Both display the same characteristic 
shape, that can be understood based on the atomic limit of the Anderson impurity model with 
$\Delta \gg \Gamma$. A straightforward extension of the $T=0$ atomic limit calculation (such as presented 
in \cite{Karrasch2008}) to the case of asymmetric level-lead couplings gives 
$\varphi_C(\epsilon)= 2 \arccos{\sqrt{g-(\epsilon/h)^2}}$. For the experimental 
$\Delta \lessapprox \Gamma$ the dependence of $g$ and $h$ on the model parameters cannot be 
trusted; we rather fit both ($g=2.2$, $h=0.72~\mathrm{meV}$) and obtain very 
good agreement. This shows that the $\epsilon$-dependence of $\varphi_C$ is a strong characteristic 
of the 0-$\pi$ transition. 

In conclusion we have 
experimentally shown that the magnetic state of a CNT quantum dot junction, 
singlet or doublet, can be controlled by the superconducting phase difference. This has to be 
contrasted to previously measured gate controlled transition. 
It is achieved by probing with unprecedented accuracy the evolution of the Josephson current at the 0-$\pi$ transition.  We have shown that the CPR has a composite behavior, with a "0" and a "$\pi$" component and that the phase at which this transition occurs is gate dependent. 
The measurements are successfully compared to exact finite temperature QMC calculations. The 
possibility to measure precisely the CPR of correlated systems motivates the study of systems 
with different symmetry such as Kondo SU(4) \cite{Zazunov2010} or with strong spin-orbit 
coupling \cite{Lim2011}.

\textit{Acknowledgments}: Ra.D., H.B. and Ri.D. thank M. Aprili, S. Autier-Laurent, J. Basset, M. Ferrier, S. Guéron, C. Li, A. Murani, and P. Simon for fruitful discussions. V.M. is grateful to D. Kennes for discussions. 

This work was supported by the French program ANR MASH (ANR-12-BS04-0016) and DYMESYS (ANR 2011-IS04-001-01).
D.J.L. also acknowledges funding by the ANR program ANR-11-ISO4-005-01 and the allocation of CPU time by GENCI (grant x2014050225).

\section{Supplementary materials}

 \subsection*{Sample fabrication}
 
\begin{figure}[htbp]
          \begin{center}
          \includegraphics[width=\columnwidth]{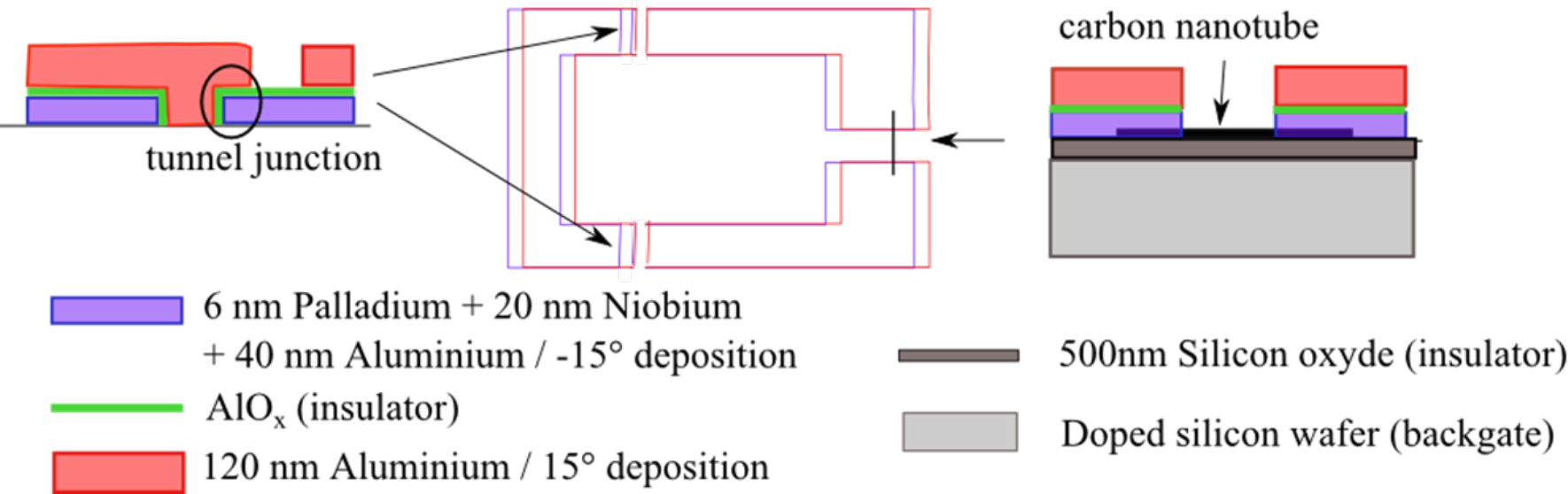}
          \end{center}
          \caption{Summary of the different layers constituting the sample}
          \label{sample}
          \end{figure}

 The carbon nanotubes (CNT) are grown by Chemical Vapor Deposition (CVD) on an oxidized doped silicon wafer \cite{Kasumov2007}. The wafer is used as a backgate. Then the three-junctions SQUID (with area $S\approx40~\mathrm{µm}^2$) is designed on top of it by electron beam lithography, with resin bridges for the tunnel junctions. A first trilayer of Pd(7~nm)/Nb(20~nm)/Al(40~nm) is deposited with an angle 15° (Fig. \ref{sample}). Its superconducting gap is found to be inhomogeneous: the CNT, which is in contact with the Pd, sees a gap $\Delta=0.17~\mathrm{meV}$ while the gap for the tunnel junction is $\Delta=0.63~\mathrm{meV}$ .
  Then the Al is oxidized and a layer of Al(120~nm) is deposited on top of it with an angle -15 ° . This layer has a superconducting gap $\Delta=0.18~\mathrm{meV}$.

 The reference JJs obtained have tunnel resistances $R_{n1}=1.6~\mathrm{k\Omega}$ and 
  $R_{n2}=1.9~\mathrm{k\Omega}$, from which we estimate the expected critical currents \cite{Ambegaokar1963}: $I_{c1}=330~\mathrm{nA}$ and $I_{c2}=280~\mathrm{nA}$. 
         
\subsection*{Additional data for the 0-$\pi$ transition}

\begin{figure}[htbp]
                  \begin{center}
                  \includegraphics[height=4.4cm]{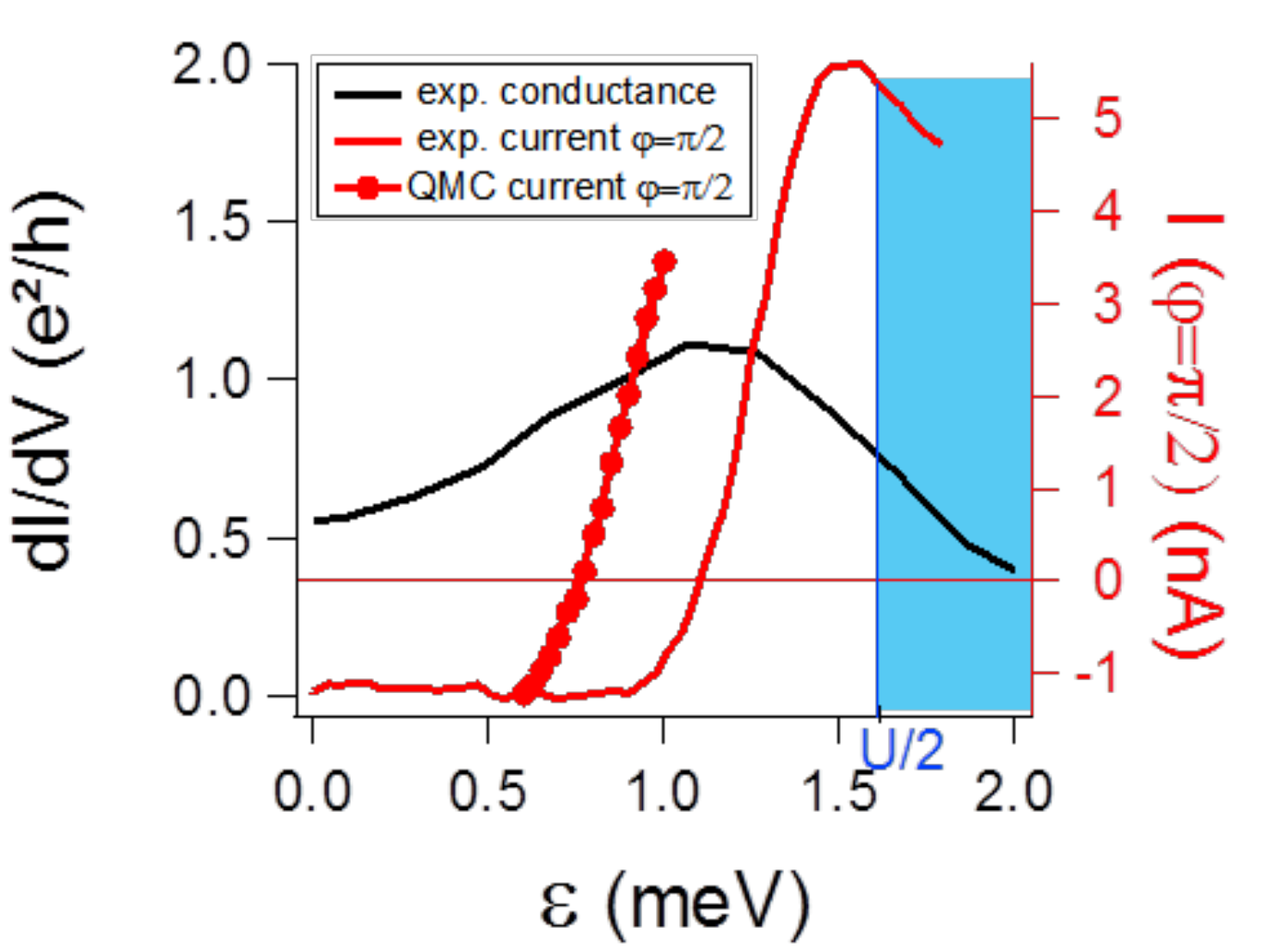}
                  \end{center}
                  \caption{Comparison of raw data for the 0-$\pi$ transition}
                  \label{transition}
                  \end{figure}

On Fig. \ref{transition} is plotted without any adjustment the supercurrent at a fixed superconducting phase difference $\varphi$ for zone A, as a function of the level energy of the dot. While the experimental data are represented with the red line, the dotted red line represents QMC data. Here, only the positive values of $\epsilon$ have been plotted.
The 0-$\pi$ transition happens for a smaller $\epsilon$ in theoretical simulation than in the measurement. 
 
\subsection*{Critical phase}

The transition from the $0$-junction to the $\pi$-junction is a first order quantum transition
and is linked to a change of the ground state from a doublet to a singlet state on the quantum dot at
a critical value of the phase $\varphi_c$. At finite but low temperatures, we only see a washed out
signal of this behavior but can nevertheless try to extract the phase boundary as a function of gate
voltage. We define the value of the critical phase at temperature $T$ by the conditions $I(\varphi_c,
T)=0$ and $\frac{\partial I}{\partial \varphi}(\varphi_c,T)<0$. At the particle-hole symmetric point 
$\epsilon=0$ and at sufficiently low temperatures $\varphi_c$ is known to only weakly dependent on 
temperature\cite{Karrasch2008} 
and will therefore provide an accurate estimate of the phase boundary. In order to check how $\varphi_c$ depends
on temperature away from the high symmetry point $\epsilon=0$, we have performed a CT-INT 
calculation at different temperatures (expressed by inverse temperatures $\beta=\frac{1}{k_B T}$) 
and for the other parameters as in the experiment.
 \begin{figure}[htbp]
        \begin{center}
        \includegraphics[height=5cm]{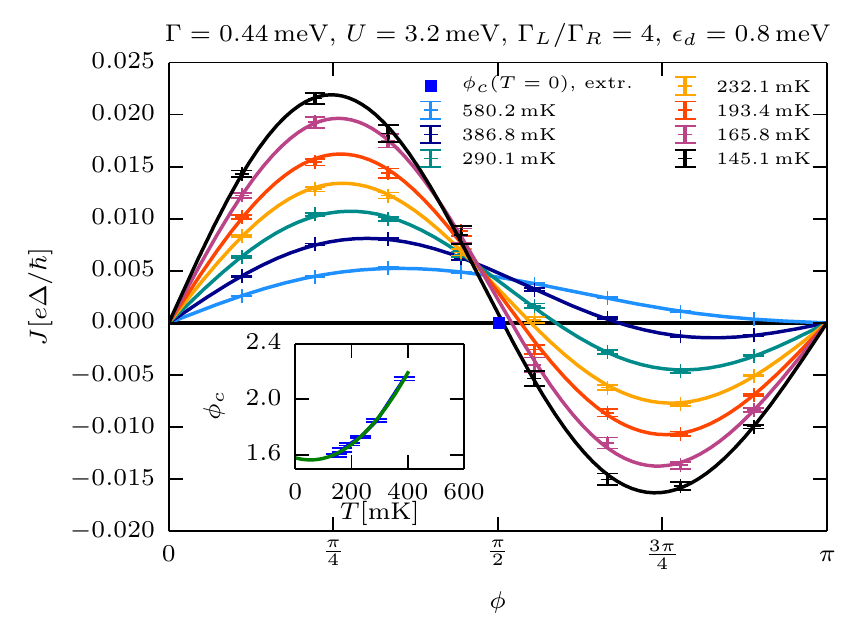}
        \end{center}
        \caption{Effect of temperature on the critical phase calculated current phase relation at 0-$\pi$ transition ($\epsilon=0.8\mathrm{meV}$) for the parameters of the experiment. The CPR has been calculated for different temperatures from 145 to 580 mK. For temperatures low enough (lower than 200 mK), the critical phase $\varphi_c$ does not depend anymore on the temperature.}
        \label{cpr_beta}
        \end{figure}
Our findings show that $\varphi_c$ indeed depends on temperatures at high temperature but
converges to the zero temperature value at low ones. The results displayed in Fig. \ref{cpr_beta}
demonstrate clearly that at the temperature $T=150\,\mathrm{mK}$ relevant to our experiment
(corresponding to $\beta\approx 77\, \frac{1}{\mathrm{meV}}$) $\varphi_c$ essentially represents the
ground state result. This shows that the result shown in the main text is in fact the accurate
ground state phase diagram.


\begin{thebibliography}{10}
\bibitem{note_EM_environment}
In particular, we noticed that the CPR at the transition can be distorted by a bad environment, acquiring a spurious even component as a function of the phase 
while the supercurrent is expected to be an odd function of the phase in absence of time-reversal symmetry breaking.
\bibitem{numerics_costs} Computing the normal state linear conductance as well as the CPRs with CT-INT is numerically 
costly. We thus focus on zone A, showing the complete transition.
\bibitem{halving_period}
 It should be noted that this $\frac{\Phi_0}{2}$ periodicity may be accidental and not fundamental since it implies that both 
the 0 and $\pi$ contributions have the same shape and amplitude which, here, is made possible by the finite temperature.
\end{thebibliography}
\end{document}